\documentclass[11pt]{article}

\usepackage[utf8]{inputenc}
\usepackage[english]{babel}

\usepackage[a4paper,margin=1in]{geometry}
\usepackage{amsmath,amssymb}

\usepackage{graphicx}
\graphicspath{{./main_figs/}{./supplement/suppl_figs/}}
\DeclareGraphicsExtensions{.pdf,.jpeg,.JPG,.png,.PNG,.eps,.tiff}
\usepackage{subcaption}
\DeclareCaptionLabelFormat{bold}{\textbf{(#2)}}
\newcommand{\labelphantom}[1]{\parbox{0pt}{\phantomsubcaption\label{#1}}}
\usepackage{booktabs}
\usepackage{multirow}
\usepackage{rotating}

\usepackage[dvipsnames]{xcolor}
\usepackage{changepage}
\usepackage{enumitem}
\usepackage{ulem}

\usepackage{authblk}

\usepackage{setspace}

\usepackage[square,numbers]{natbib}

\usepackage[colorlinks=true, linkcolor=black, citecolor=black, urlcolor=blue, filecolor=blue, breaklinks=true]{hyperref}

\usepackage[capitalise,noabbrev,nameinlink]{cleveref}
\crefdefaultlabelformat{#2\textbf{#1}#3}
\Crefname{figure}{\textbf{Figure}}{\textbf{Figures}}
\Crefname{section}{\textbf{Section}}{\textbf{Sections}}
\Crefname{table}{\textbf{Table}}{\textbf{Tables}}

\usepackage{pgffor}
\usepackage{alphalph}

\usepackage{xspace}                

\newcommand{\generateFigSubpanels}[6][0]{
     \expandafter\newcommand\csname#2\endcsname{ 
      \begin{figure}[htbp]
        \foreach [count=\i] \x in #6{%
            \labelphantom{fig:#2:\AlphAlph{\i}}
        }
        \centering
        \includegraphics[width=#4\linewidth,angle=#1]{#3}
        \caption{\textbf{\normalsize #5}}\label{fig:#2}
        \footnotesize
        \justifying
        \foreach [count=\i] \x in #6{%
            \noindent\subref{fig:#2:\AlphAlph{\i}}~\x\space
        }
      \end{figure}
    }
}

\newcommand{\generateFig}[6][0]{
     \expandafter\newcommand\csname#2\endcsname{ 
      \begin{figure}[htbp]
        \centering
        \includegraphics[width=#4\linewidth,angle=#1]{#3}
        \caption{\textbf{\normalsize #5}\label{fig:#2}
        \footnotesize
        \normalfont
        #6
        }
        
      \end{figure}
    }
}

\newcommand{\generateSidewaysFigSubpanels}[6][0]{
     \expandafter\newcommand\csname#2\endcsname{ 
      \begin{sidewaysfigure}[htbp]
        \foreach [count=\i] \x in #6{%
            \labelphantom{fig:#2:\AlphAlph{\i}}
        }
        \centering
        \includegraphics[width=#4\linewidth,angle=#1]{#3}
        \caption{\textbf{\normalsize #5}}\label{fig:#2}
        \footnotesize
        \justifying
        \foreach [count=\i] \x in #6{%
            \noindent\subref{fig:#2:\AlphAlph{\i}}~\x\space
        }
      \end{sidewaysfigure}
    }
}

\newcommand{\generateSidewaysFig}[6][0]{
     \expandafter\newcommand\csname#2\endcsname{ 
      \begin{sidewaysfigure}[htbp]
        \centering
        \includegraphics[width=#4\linewidth,angle=#1]{#3}
        \caption{\textbf{\normalsize #5}\label{fig:#2}
        \footnotesize
        \normalfont
        #6
        }
        
      \end{sidewaysfigure}
    }
}

\newcommand{\generateTab}[6][]{
     \expandafter\newcommand\csname#2\endcsname{ \begin{table}[ht]
          \centering
                \resizebox{#4\textwidth}{!}{
                \input{#3}
            }
            \caption{\textbf{\normalsize #5}\label{tab:#2}
            \footnotesize
            \normalfont
            #6
            }
            
        \end{table}
    }
}

\newcommand{\generateSidewaysTab}[6][]{
     \expandafter\newcommand\csname#2\endcsname{ \begin{sidewaystable}[ht]
          \centering
                \resizebox{#4\textwidth}{!}{
                \input{#3}
            }
            \caption{\textbf{\normalsize #5}\label{tab:#2}
            \footnotesize
            \normalfont
            #6
            }
            
        \end{sidewaystable}
    }
}

\begin{document}



\title{Evaluating DNA function understanding in genomic language models using evolutionarily implausible sequences}


\author[1,2,*]{Shiyu~Jiang}
\author[1]{Xuyin~Liu}
\author[1,*]{Zitong~Jerry~Wang}
\affil[1]{Center for Interdisciplinary Studies, School of Science, Westlake University, Hangzhou, China}
\affil[2]{Department of Quantitative and Computational Biology, University of Southern California, Los Angeles, CA, United States}

\affil[ ]{ } 
\affil[*]{\textit{Correspondence to }\underline{shiyujia@usc.edu} and \underline{jerry@westlake.edu.cn}}
\renewcommand\Affilfont{\footnotesize} 
\setcounter{Maxaffil}{0} 
\date{} 


\newcommand{\makeAbstract}{
\begin{abstract}
Genomic language models (gLMs) hold promise for generating novel, functional DNA sequences for synthetic biology. However, realizing this potential requires models to go beyond evolutionary plausibility and understand how DNA sequence encodes gene expression and regulation. We introduce a benchmark called Nullsettes, which assesses how well models can predict \textit{in silico} loss-of-function (LOF) mutations, in synthetic expression cassettes with little evolutionary precedent. Testing 12 state-of-the-art gLMs, we find that most fail to consistently detect these strong LOF mutations. All models show a sharp drop in predictive accuracy as the likelihood assigned to the original (nonmutant) sequence decreases, suggesting that gLMs rely heavily on pattern-matching to their evolutionary prior rather than on any mechanistic understanding of gene expression. Our findings highlight fundamental limitations in how gLMs generalize to engineered, non-natural sequences, and underscore the need for benchmarks and modeling strategies that prioritize functional understanding.\newline

\noindent{\textbf{Keywords:} genomic language model, synthetic biology, mutation prediction, gene expression, regulatory genomics}

\end{abstract}

}

\setstretch{1.15}

\maketitle
\makeAbstract
\clearpage

\section{Background}

\noindent{Genomic language models (gLMs) learn a probability distribution over naturally occurring DNA sequences, approximating the evolutionary plausibility of a sequence or its probability of arising and persisting under evolutionary constraints~\cite{benegas2025genomic, consens2025transformers}. This plausibility reflects factors such as the presence of common regulatory motifs or similarity to known natural genomes. For example, gLMs tend to assign higher likelihoods to promoters with frequent natural motifs than to random synthetic promoters.}

Evolutionary plausibility can be a useful proxy for biological function, as gLMs have already demonstrated strong zero-shot performance in predicting mutation effects to support sequence design. In both genomic and protein language models (gLMs and pLMs), higher model likelihoods often correlate with improved functional properties. For example, pLM likelihoods can guide the design of higher 
antibodies~\cite{hie2024efficient}, more efficient base editors~\cite{he2024protein}, brighter fluorescent proteins~\cite{zhang2025integrating}, and deep mutational scaning (DMS) tasks~\cite{su2024saprothub,notin2023proteingym}. Similarly, gLM likelihoods enable genome-wide prediction of variant effects~\cite{GPN} and 5$'$ untranslated region (UTR) optimization~\cite{chu20245}.

However, it remains unclear whether gLMs trained solely on natural genomes understand DNA function well enough to generalize to synthetic constructs with little to no evolutionary precedent. Current evaluations for gLMs on mutation effect prediction focus primarily on natural sequences, such as variants within endogenous cis-regulatory elements and splice sites~\cite{tang2025evaluating}. In contrast, synthetic biology often requires the design of functional sequences with little evolutionary history, to solve industrial challenges, avoid crosstalk with native pathways, or push expression levels beyond natural limits. Examples include ultra-strong synthetic promoters~\cite{schlabach2010synthetic}, miRNA-based regulatory circuits that achieve dosage-compensated gene expression using elements orthogonal to native miRNAs~\cite{du2024mirna}, and engineered metabolic pathways for production of small molecule drugs~\cite{yan2023applications}. We need evaluation benchmarks that test whether models can generalize beyond evolutionary patterns to capture functional principles to help rewire biology.

Fundamental mechanisms of gene expression are conserved across all domains of life, making it an ideal test of functional understanding. We introduce, Nullsettes, a benchmark for evaluating whether gLMs can predict loss-of-function (LOF) mutations in synthetic expression cassettes. We create these LOF mutations in silico by rearranging key regulatory elements, such as promoters and start codons, within nonmutant cassettes. These rearrangements disrupt the canonical sequence structure required for transcription and translation, rendering the sequences non-functional. We curate nonmutant cassettes from massive parallel reporter assay (MPRA) datasets ~\cite{Lagator_data, deBoer_data, Kosuri_data, Zahm_data}. GLMs assign low likelihood to these cassette as they contain genetic elements from evolutionarily distant species, and incorporate random sequences functioning as promoters. 

Across 12 state-of-the-art models (and 28 variants) spanning a wide range of design choices, we find that gLMs perform poorly on zero-shot LOF prediction, particularly under contexts that deviate strongly from natural sequence statistics. For example, model performance worsens when mutations cause more severe functional disruption, or when cassettes include strong promoters drawn from random sequence libraries. Notably, all models show a sharp drop in prediction accuracy as the likelihood of the nonmutant sequence decreases, indicating that gLMs rely heavily on pattern-matching to their evolutionary priors rather than on functional understanding of gene expression. Accurate predictions only emerge when the nonmutant likelihood surpasses a length-scaled threshold. Comparisons between models show that functional generalization may depend more on the relevance and quality of pretraining data than on the sheer size of the dataset. These findings highlight fundamental limitations in how current gLMs generalize to engineered sequences, and underscore the need for benchmarks and modeling strategies that prioritize the functional understanding of DNA sequences.








\generateFigSubpanels{figureSuggestions}{figureSuggestions}{1}
    {Making good figures requires attention to detail.} 
    {{
        {General principals of making and labelling a good figure. These should help facilitate understanding and reproducibility. }, 
        {Some more detailed technical specifics about making figures and their sizes. Inkscape is a great free vector figure editor and alternative to Adobe Illustrator.\captionTips} 
    }}

\section{Results and discussion}

\begin{figure*}[!ht]
    \centering
    \includegraphics[width=\linewidth]{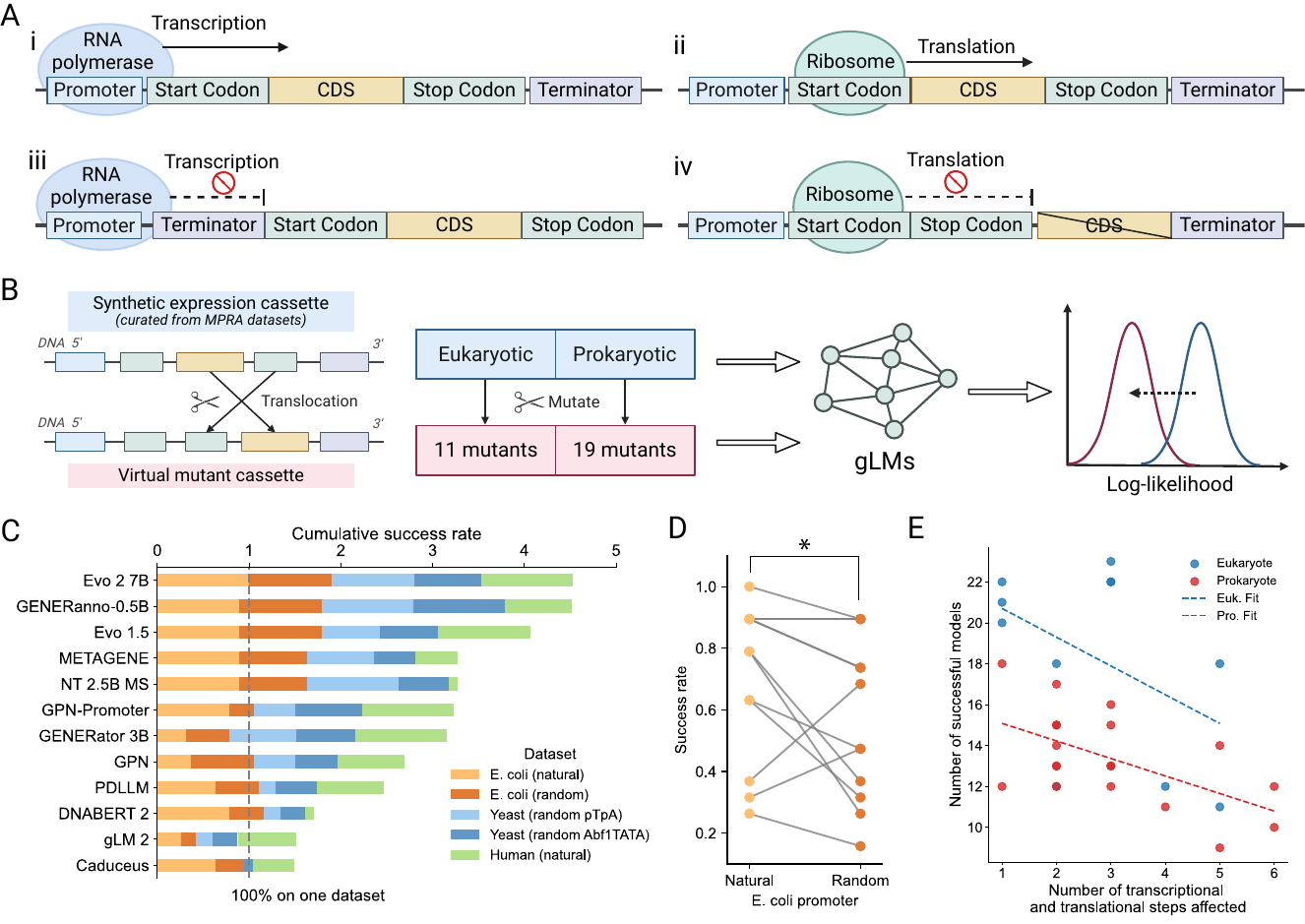}
    \caption{
        \textbf{Overview of Nullsettes.} A) Schematic examples showing transcription (i) and translation (ii) becoming impaired when the order of control element such as terminator (iii) or stop codon (iv) are altered. B) Nullsettes consist of 11 and 19 mutant variants for each eukaryotic and prokaryotic expression cassette, respectively, by translocating control elements, and evaluates how well gLMs can predict these loss-of-function mutations based on changes in log-likelihood. We curate functional nonmutant cassettes from five MPRA datasets. C) Cumulative success rates of representative gLMs evaluated on five MPRA datasets, with promoter either being random sequences or based on natural promoter (indicated in bracket). The two yeast datasets use promoter containing both natural motifs (Abf1-TATA, polyT-polyA) and random sequences. Dashed lines denote 100\% success on one dataset. Models are sorted by cumulative success rate. 
        D) Success rates of models in identifying Nullsettes mutants for \textit{E. coli} expression cassettes with either naturally-derived or random promoters. Each line connects the performance of a single model across the two sets of cassettes. Statistical significance with $p<0.05$ for a one-sided paired T-test.
        E) Scatter plot depicting the relationship between the number of transcriptional and translational steps affected (x-axis) and the number of models that successfully assign low pseudo-likelihood (y-axis) across Nullsettes mutants. Each point represents a distinct type of Nullsettes mutant, with eukaryotic (n = 11) and prokaryotic (n = 19) variants shown in blue and red, respectively. Dashed lines indicate linear regression fits for each group.
    }
    \label{fig:fig1}
\end{figure*}

We design Nullsettes as a benchmark to test whether genomic language models (gLMs) understand functional gene expression in sequences that lack evolutionary plausibility. Nullsettes consists of zero-shot mutation effect prediction tasks, where synthetic expression cassettes are subjected to virtual loss-of-function (LOF) mutations created by rearranging key genetic elements.

In a canonical eukaryotic expression cassette (Figure~\ref{fig:fig1}A, i–ii), control elements including the promoter, start codon, coding sequence (CDS), stop codon, and terminator must appear in a specific 5$'$-3$'$ order to enable proper transcription and translation. Rearranging these elements, for example, placing a terminator upstream of the CDS (iii) or positioning the CDS downstream of the stop codon (iv), disrupts this structure and results in nonfunctional gene expression. Nullsettes focuses on single-element translocations that eliminate functional protein expression and cannot be rescued by any upstream or downstream sequences (Figure~\ref{fig:fig1}B). From these criteria, we obtain 11 mutation types for eukaryotic and 19 for prokaryotic cassettes, with additional prokaryotic variants involving rearrangement of ribosome binding site (RBS) (Supplementary Table~\ref{tab:prok_virtual_mutant},~\ref{tab:euk_virtual_mutant}).

To ensure these tasks test generalization beyond evolutionary prior, we generate virtual mutants from synthetic expression cassettes with low gLM-assigned likelihoods but strong gene expression in either \textit{E. coli}, yeast, or human, drawn from five massively parallel reporter assay (MPRA) datasets~\cite{Lagator_data, deBoer_data, Kosuri_data, Zahm_data} (Figure~\ref{fig:fig1}B). These cassettes combine sequences from distantly related species, for example, GFP as CDS from \textit{Aequorea victoria} with promoters and terminators from bacteria, human, or mouse. Many cassettes even use random (but functional) promoters, which further lower gLM-assigned likelihood (Supplementary Figure~\ref{SIfig:fig1}). Together, these cassettes enable evaluation of model generalization to functional yet evolutionarily implausible sequences across species.

Given all cassettes in a dataset, we evaluate mutation effect prediction in a zero-shot setting by comparing the distribution of nonmutant log-likelihood (LL) to the distribution of corresponding Nullsettes mutants (Figure~\ref{fig:fig1}B). We consider a model capable of identifying a mutation type if the mutant LL are significantly lower than the nonmutant LL, as determined by a paired t-test. A model’s success rate on the dataset is the proportion of mutation types identified.

\begin{figure*}[!ht]
    \centering
    \includegraphics[width=\linewidth]{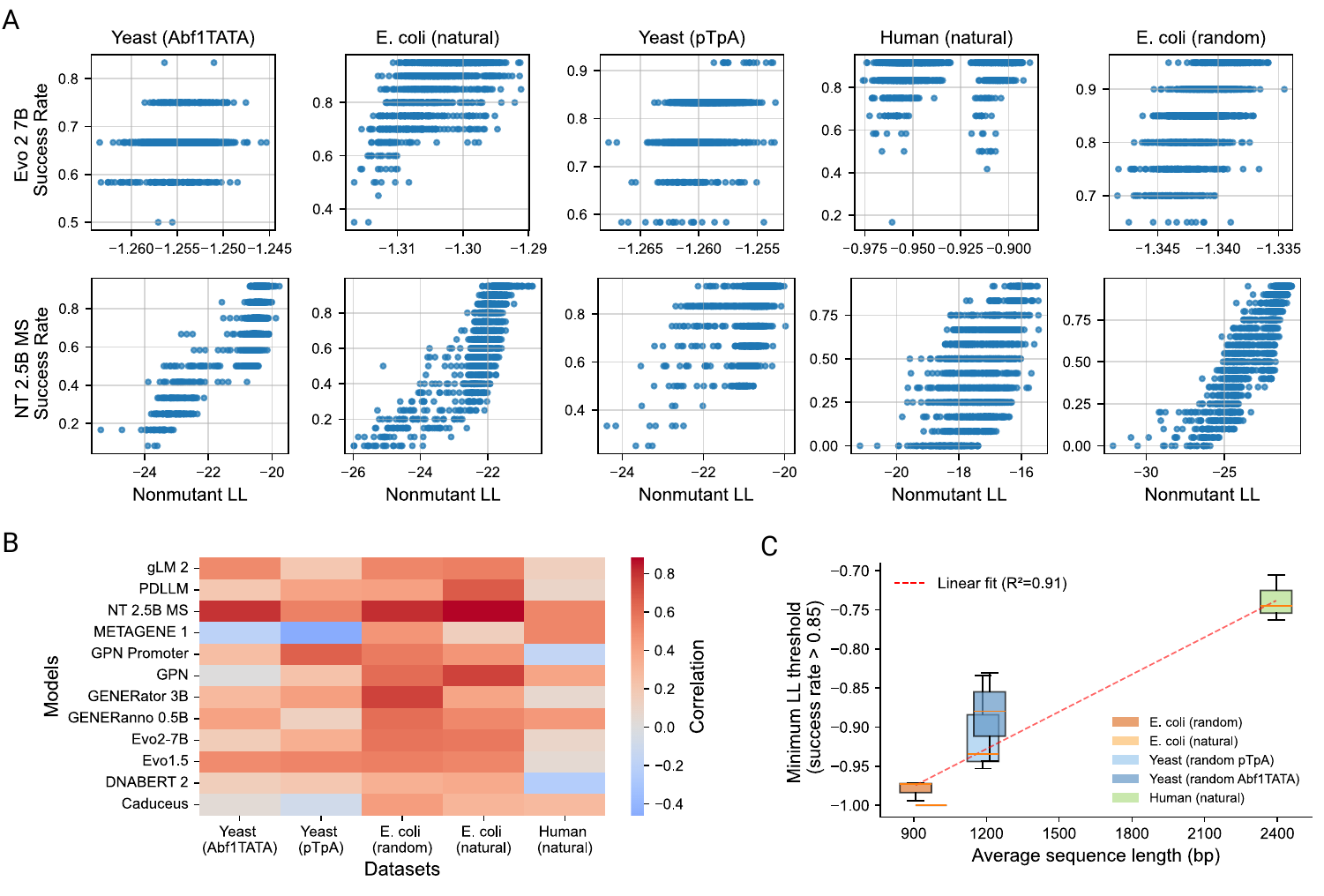}
    \caption{\textbf{Relationship between gLM likelihood and zero-shot performance on mutation effect prediction.} A) Each point represents an expression cassette. Scatterplots show the Nullsettes prediction performance of Evo2-7B and NT-2.5B-MS for cassettes from five datasets. B) Heatmap showing correlation between the gLM log likelihood (LL) of a sequence and gLM success rate on predicting its corresponding Nullsettes, across five datasets and 12 models. 
    C) Log-likelihood (LL) threshold to achieve a success rate $>0.85$ across datasets of increasing sequence length. Boxplots show the distribution of LL thresholds for top performing models (GENERanno, Evo2-7B, and Evo1.5) across five datasets, rescaled by lowest threshold. Datasets are ordered according to their average sequence length. Red dashed line is linear regression fit to the median threshold for each dataset.}
    \label{fig:fig2}
\end{figure*}

Across five datasets, we found that most genomic language models perform poorly in predicting LOF mutants in Nullsettes. We evaluated 12 representative models (and 28 variants; Supplementary Table~\ref{tab:rep_model},~\ref{glm_detail_clm}, and~\ref{glm_detail_mlm}) spanning state-of-the-art approaches to genomic language modeling. Nine out of twelve models achieved $<50\%$ success rate on at least one dataset, failing to identify strong LOF mutations (Figure~\ref{fig:fig1}C). Only Evo2-7B and GENERanno-0.5B deliver strong, consistent performance (Supplementary Figure~\ref{SIfig:fig2}). GENERanno matched Evo2-7B performance despite having 14-fold fewer parameters and using 12-fold less pretraining data, potentially due to its curated pretraining on actively expressed regions of the genome~\cite{GENERanno}, suggesting that data quality and relevance may matter more than scale for functional generalization. Supporting this hypothesis, performance did not consistently improve with scaling across four model families (Supplemental Figure~\ref{SIfig:fig3}).

Current gLMs recognize what natural expression cassettes look like, but not how they work, failing to generalize to contexts that deviate strongly from natural sequence statistics. Across most models, performance drops significantly when identifying mutants with strong promoters drawn from random sequence libraries compared to promoters derived from nature (\textit{p} $= 0.04$)(Figure~\ref{fig:fig1}D). When evolutionarily plausibility is not a useful prior for function, as these random sequences are assigned significantly lower LL (Supplementary Figure~\ref{SIfig:fig1}), models struggle to predict mutation effects.

A similar failure emerges across Nullsette mutations, with model performance worsening as mutations disrupt more steps of the gene expression process (Figure~\ref{fig:fig1}E). If models understood how genetic elements mechanistically drive expression, their predictions should remain robust or even improve as mutations become more disruptive. Instead, performance declines, suggesting that models do not reason about functional mechanism, and behave sensibly only within the distribution of natural sequences. We quantify disruption by counting the transcribed and translated elements missing in each mutant. When only one step is affected (e.g., a premature stop codon), the mutant remains close to natural sequence space, and models can often detect the resulting loss of function. But when multiple steps are disrupted (e.g., loss of both transcription and translation), the sequence falls further outside the distribution seen during training, and model outputs become unreliable. Taken together, these results show that current gLMs do not understand regulatory logic, generalizing based on sequence similarity to evolutionary prior.

Across all gLMs, zero-shot performance in mutation effect prediction declines sharply as the log-likelihood (LL) of the original (nonmutant) sequence decreases, reinforcing that models rely on their evolutionary priors rather than a mechanistic understanding of DNA function.
If a model understood how mutations disrupt gene expression, its predictions should depend on what function the mutation alters, not on how similar the sequence looks to the model’s training corpus. In that case, LOF mutants would be correctly identified regardless of the LL of the nonmutant cassette. However, we do observe this dependency on LL.
Figure~\ref{fig:fig2}A shows this relationship for Evo2-7B and NT-2.5B-MS.
For both models, the ability to detect LOF mutations improves consistently as the LL of the nonmutant increases. 
For example, in the E. coli (random) dataset~\cite{Lagator_data}, NT-2.5B-MS fails almost completely when LL $< -27$, but performs well above $80\%$ when LL $> -22$.
This dependency is not specific to a single model. As shown in Figure~\ref{fig:fig2}B, the same correlation between nonmutant LL and prediction accuracy holds across nearly all models, regardless of architecture or training corpus, indicating a shared reliance on sequence prior rather than reasoning about expression mechanisms.

Among top-performing models, the threshold LL required for high prediction accuracy is consistent and scales linearly with sequence length. Figure~\ref{fig:fig2}A shows that the minimum LL threshold for accurate prediction can vary widely across datasets. For example, NT-2.5B-MS achieves a success rate of $0.85$ at a threshold LL of approximately $–20$ for Yeast (Abf1TATA) cassettes, but around $–16$ for Human (natural). For each model–dataset pair, we estimate the LL threshold corresponding to a success rate of $0.85$ using linear regression. Figure~\ref{fig:fig2}C reveals that these thresholds are consistent among top performing models (after rescaling by lowest threshold). Furthermore, this threshold scales linearly with sequence length, showing a simple linear model can predict an effective LL threshold for zero-shot mutation effect prediction.

\section{Conclusions}

In this study, we introduce Nullsettes, a benchmark for evaluating the ability of genomic language models (gLMs) to understand DNA function by predicting loss-of-function (LOF) mutations in synthetic expression cassettes. Our analysis of 12 state-of-the-art models reveals that most struggle to consistently identify these LOF mutations, often relying on surface-level similarity rather than true functional understanding of regulatory elements. Reinforcing this fact, we show that LOF prediction accuracy declines sharply when nonmutant sequences have low gLM-assigned log-likelihood, as is the case for synthetic sequences. Comparing performance between models, we find a key avenue for improving functional understanding may be prioritizing curated, functionally relevant pretraining data over brute-force scaling. Together, these findings expose critical limitations in current gLMs and highlights the need for benchmark and modeling strategies that prioritize functional understanding for cell engineering.
\section{Methods}
\subsection{Functional cassette curation}
\subsubsection{Prokaryote}
To evaluate the regulatory grammar of prokaryotes on genomic language models, we utilized two benchmark datasets derived from synthetic expression constructs in \textit{Escherichia coli}. The first, the Kosuri dataset \cite{Kosuri_data}, comprises combinatorial assemblies of promoters and ribosome binding sites (RBS) upstream of a superfolder GFP (sfGFP) reporter, designed to dissect transcriptional and translational contributions to gene expression. The second, the Lagator dataset \cite{Lagator_data}, features a large-scale library of random promoter sequences with experimentally measured activity, enabling evaluation of models’ ability to generalize to highly diverse, out-of-distribution regulatory inputs. 
\subsubsection{Eukaryote}
To evaluate the capacity of genomic language models to generalize to eukaryotic regulatory logic, we utilized two large-scale massively parallel reporter assay (MPRA) datasets. The Zahm dataset \cite{Zahm_data} comprises a library of 6,144 synthetic promoters constructed by combining transcriptional response elements (TREs) from 229 human and mouse transcription factors with minimal promoters. These constructs were assayed across multiple human cell lines and stimulatory conditions to quantify dynamic, stimulus-specific gene expression. The deBoer dataset \cite{deBoer_data} includes a comprehensive collection of 100 million randomized promoter sequences in yeast, enabling high-resolution dissection of cis-regulatory grammar under out-of-distribution scenario.

\subsubsection{Promoter selection}
To construct decoupled expression cassettes, we selected 1,500 promoters or promoter–RBS pairs from four distinct datasets.

For the Kosuri dataset~\cite{Kosuri_data}, we selected candidate promoter–RBS pairs with protein expression levels exceeding $\mu + 1.5\sigma$ (10,165) across the full distribution. In contrast, given the limited number of highly active constructs in the Lagator dataset, we directly selected the top 1,500 promoter sequences ranked by protein output. We then compared the log-likelihood (LL) distributions of promoter–RBS groups using genomic language models including Evo1-8k, GENERator-3B, GPN, and NT-2.5B-MS. As the Lagator dataset~\cite{Lagator_data} consists of randomly generated promoters, whereas the Kosuri library is more rationally designed, we prioritized Kosuri promoter–RBS pairs whose LL distributions surpassed those from Lagator (see Supplementary Figure~\ref{SIfig:fig1}A), yielding 1,500 expression cassettes from each dataset.

deBoer dataset~\cite{deBoer_data} contains two promoter libraries: Abf1TATA and pTpA. Abf1TATA is designed by embedding conserved transcription factor binding sites such as Abf1 and a canonical TATA box, mimicking features of natural yeast promoters. In addition, pTpA consists of synthetic promoters constructed with a poly-T–poly-A architecture. Poly(dA-dT) sequences are naturally occurring motifs that associate with the centers of nucleosome-free regions in yeast promoters and serving as a hybrid natural + randomized library. We selectively constructed 1,500 active promoters for each dataset .

The Zahm dataset \cite{Zahm_data} comprises synthetic promoters built by coupling one of three minimal promoters—minCMV, minProm, or minTK—with diverse transcriptional response element (TRE) units. TRE units contain Transcription factor binding motifs (TFBMs) mostly derived from human genomes. To construct a representative set of mammalian expression cassettes, we selected the top 1,500 promoter-TRE combinations exhibiting the highest transcriptional activity.

\subsection{Nullsettes Virtual Mutant Construction}
To assess whether genomic language models can detect syntactic violations in gene regulatory architecture, we constructed Nullsettes—a benchmark of completely non-functional expression cassettes. These sequences were generated by systematic component translocations that disrupt the canonical transcription–translation structure while retaining all original components.

We first defined the canonical regulatory architecture for both prokaryote and eukaryote:

\textbf{Prokaryotic systems} (e.g., \textit{E. coli}) contain six ordered elements: \texttt{Promoter - RBS - Start codon - CDS - Stop codon - Terminator}.

\textbf{Eukaryotic systems} (e.g., yeast, mammalian) consist of five ordered elements, lacking a RBS as translation initiation is guided by the 5$'$ cap and the Kozak consensus sequence rather than a defined RBS: \texttt{Promoter - Start codon - CDS - Stop codon - Terminator}.

For each system, we generated all possible single-component translocation mutants, in which one element was relocated to a different position while preserving the identity and total count of elements.

To determine whether a mutant retained functional syntax, we evaluated each sequence against a minimal set of biologically grounded ordering constraints. These constraints ensure that proper CDS expression cannot be rescued by any flanking sequences outside the cassette. Specifically, we consider all circular permutations of each ordering and retain a mutant as a Nullsette only if no circular permutation of the mutant satisfies all three of the following rules:

\vspace{1ex}
\textbf{Rule 1: Proper initiation ordering} — Ensures that transcription starts upstream of translation initiation and that the CDS is translated from an upstream start codon.\\
Eukaryotic: \quad Position(Promoter) $<$ Position(Start codon) $<$ Position(CDS)\\
Prokaryotic: \quad Position(Promoter) $<$ Position(RBS) $<$ Position(Start codon) $<$ Position(CDS)

\vspace{1ex}
\textbf{Rule 2: Valid translation termination} — The stop codon must follow the CDS or precede the start codon in circular order.\\
(i) Position(CDS) $<$ Position(Stop codon) \quad or \quad (ii) Position(Stop codon) $<$ Position(Start codon)

\vspace{1ex}
\textbf{Rule 3: Valid transcription termination} — The terminator must lie downstream of the CDS or upstream of the promoter in circular form.\\
(i) Position(CDS) $<$ Position(Terminator) \quad or \quad (ii) Position(Terminator) $<$ Position(Promoter)

\vspace{1ex}
After applying these rules across all circular permutations, we identified 19 non-functional translocation mutants in the prokaryotic system and 11 in the eukaryotic system, all of which violated the defined minimal functional syntax. These mutants are synthetically designed to be completely non-functional, challenging models to detect biologically catastrophic yet syntactically subtle disruptions in regulatory element ordering.

All mutant sequences are listed in Supplementary Table~\ref{tab:prok_virtual_mutant} (prokaryote) and Supplementary Table~\ref{tab:euk_virtual_mutant} (eukaryote).

\subsection{Baseline Models}
We benchmarked a diverse set of 12 self-supervised genomic foundation models that represent current state-of-the-art approaches to DNA language modeling. 
In general, the models can be categorized based on their tokenization schemes (fixed-length k-mers, single-nucleotide tokens, byte-pair encoding, and hybrid schemes), pretraining strategies (masked language modeling \cite{mask_lang_model} vs. autoregressive modeling \cite{next_token_prediction}), training corpus diversity (human references, plant genomes, multispecies genomes, and large-scale metagenomic assemblies), and architectural paradigms (CNN-based models, Transformer-based models \cite{consens2025transformers}, and emerging architectures like StripedHyena \cite{stripedhyena,ku2025systems} and Mamba \cite{gu2023mamba}). Detailed model specifications are concluded in Supplementary Table \ref{glm_detail_clm} for casual language modeling-based models and \ref{glm_detail_mlm} for mask language modeling-based models.

\subsection{Evaluation Metrics}
We computed the mean base-pair log-likelihood (LL) as the sequence-level LL score. To account for the distinct pretraining objectives of causal language models and masked language models, we applied different LL computation strategies.

Specifically, for casual language models (CLM), the log-likelihood of a nucleic acid sequence $X=(x_1, x_2, ..., x_n)$ is computed using logits output from the model. Since CLM operates in an autogressive manner, each token $x_t$ is predicted based on all previous token $x<t$. Given the model's ${logits}_t$ at each position $t$, the probability of the ground-truth token is obtained via softmax:
\begin{equation}
P(x_t \mid x_{<t}) = \frac{\exp(\text{logits}_{t, x_t})}{\sum_{v \in V} \exp(\text{logits}_{t, v})}
\end{equation}
where $V$ represents the vocabulary. The mean log-likelihood is then computed as:
\begin{equation}
\text{LL}_{\text{CLM}}(X) = \frac{1}{n-1} \sum_{t=2}^{n} \log P(x_t \mid x_{<t})
\end{equation}

Masked language models (MLMs) differ from autoregressive models in their inference behavior, as they are not inherently designed for sequential generation. To address this, \citet{wang2019bert} introduced the pseudo log-likelihood (PLL) approach, wherein each token in a sequence is masked individually to compute token-wise conditional probabilities. More recently, \citet{gordon2024protein} proposed Single-Inference Pseudo Log-Likelihood, an efficient approximation that enables linear-time inference for BERT-style MLMs by exploiting training-time masking dynamics.

Rather than computing pseudo log-likelihood (PLL), we assessed relative changes in mean token-wise log-probabilities derived from unmasked logits for all MLM-based gLMs. This method, though lacking strict probabilistic interpretation, provides a scalable and effective proxy for quantifying model sensitivity to syntactic or functional disruptions. The formula is as follows:

Given the ${logits}_t$ at each position $t$, the probability of the correct token is computed as:
\begin{equation}
P(x_t \mid X) = \frac{\exp(\text{logits}_{t, x_t})}{\sum_{v \in V} \exp(\text{logits}_{t, v})}
\end{equation}
Then the log-likelihood for the entire sequence is computed over all positions:
\begin{equation}
\text{LL}_{\text{MLM}}(X) = \frac{1}{n} \sum_{t=1}^{n} \log P(x_t \mid X)
\end{equation}
where all positions contribute to the likelihood computation since no masking is performed during inference.

To assess significance between paired conditions (e.g., original vs. mutant), we performed as one-sided paired permutation test. Given differences $d_i = {x_i}^{mutant} - {x_i}^{original}$ for each pair $i$, we computed the observed mean $\bar{d}_{\text{obs}}$. Under the null hypothesis of no effect, signs of $d_i$ are exchangeable. We generated $N=10,000$ random permutations by sampling $s_i \in \{-1, 1\}$ and computing $\bar{d}^{(j)} = \frac{1}{n} \sum_{i=1}^{n} s_i d_i$. The one-sided p-value was calculated as:
\begin{equation}
p = \frac{1}{N} \sum_{j=1}^{N} \mathbb{I} \left( \bar{d}^{(j)} \leq \bar{d}_{\text{obs}} \right)
\end{equation}
for the alternative hypothesis that the mutant is smaller than the original.

\section{Data availability}
The expression cassette of Kosuri dataset~\cite{Kosuri_data} was obtained from \url{https://www.addgene.org/47441/}. The Lagator dataset~\cite{Lagator_data} was downloaded from \url{https://github.com/szarma/Thermoters}. The deBoer dataset~\cite{deBoer_data} was downloaded from \url{https://www.ncbi.nlm.nih.gov/geo/query/acc.cgi?acc=GSE104878}. The Zahm dataset~\cite{Zahm_data} was downloaded from \url{https://www.ncbi.nlm.nih.gov/geo/query/acc.cgi?acc=GSE271608}. 

\section{Code availability}
The source code for benchmarking gLMs and the Nullsettes used in this work is publicly available at: \url{https://github.com/cellethology/GLM-Nullsette-Benchmark}.

\section{Acknowledgements}
The authors thank members of the Cell Ethology Lab for valuable comments and suggestions. 
This work was supported by the Westlake Fellows Program at Westlake University.

\section{Author Contributions}
Z.J.W. conceived the initial proposal. S.J. and Z.J.W. jointly developed the overall concept and contributed to the framework design. S.J. implemented the framework. S.J., X.L., and Z.J.W. conducted and analyzed the computational experiments. S.J. and Z.J.W. wrote the initial draft of the manuscript.


\section{Competing interests}

The authors declare no competing interests.

\begin{singlespace}
  \bibliographystyle{unsrtnat}  
  \bibliography{library}        
\end{singlespace}

\clearpage
\section*{Supplementary Information}
\setcounter{page}{1}
\subsection*{Supplemental Figures}
\setcounter{figure}{0} 
\renewcommand{\thefigure}{S\arabic{figure}} 

\begin{figure*}[!ht]
    \centering
    \includegraphics[width=0.9\linewidth]{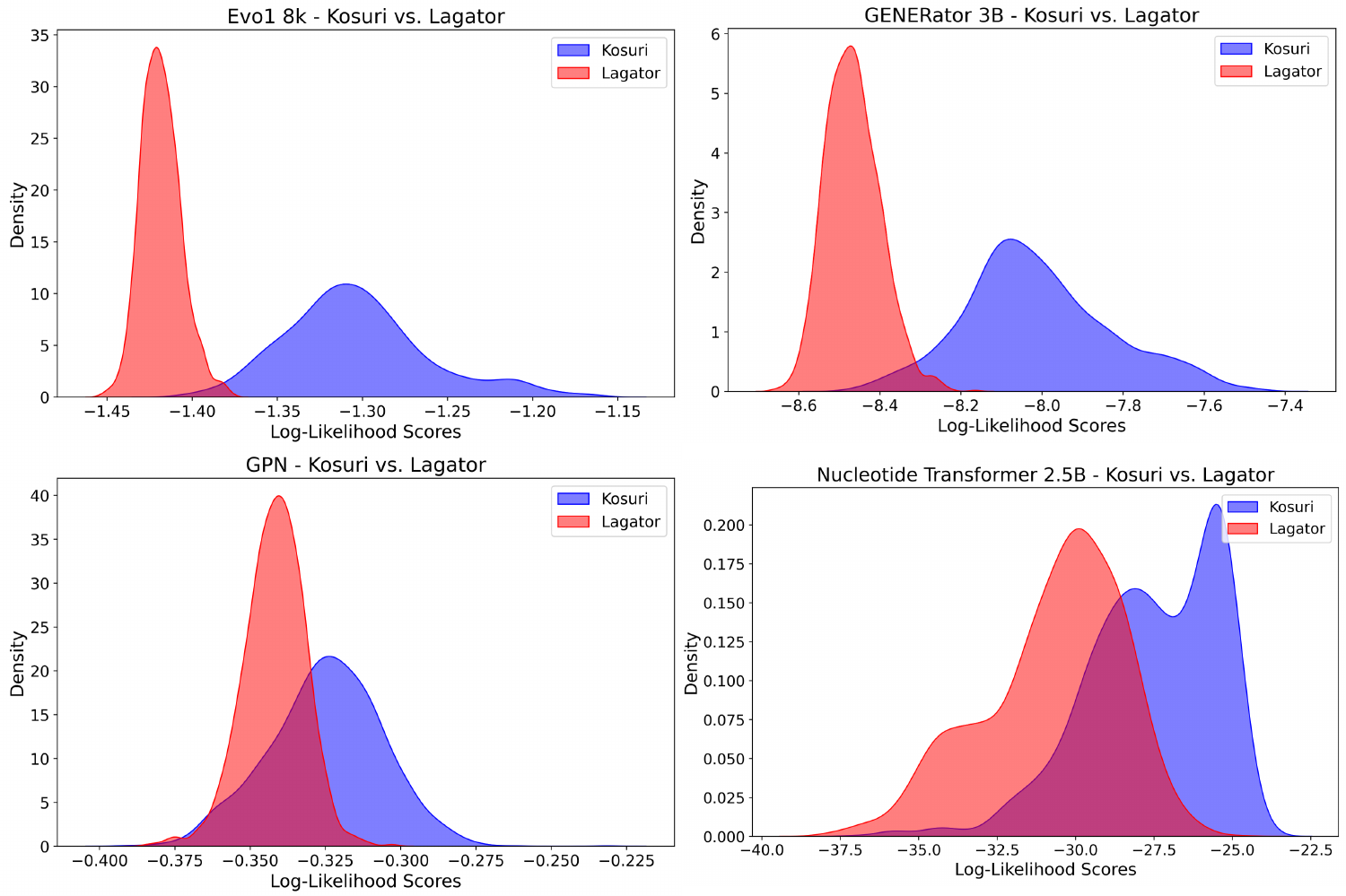}
    \caption{
        \textbf{Comparison of promoter log-likelihood distribution}
    }
    \label{SIfig:fig1}
\end{figure*}


\begin{figure*}[!ht]
    \centering
    \includegraphics[width=0.9\linewidth]{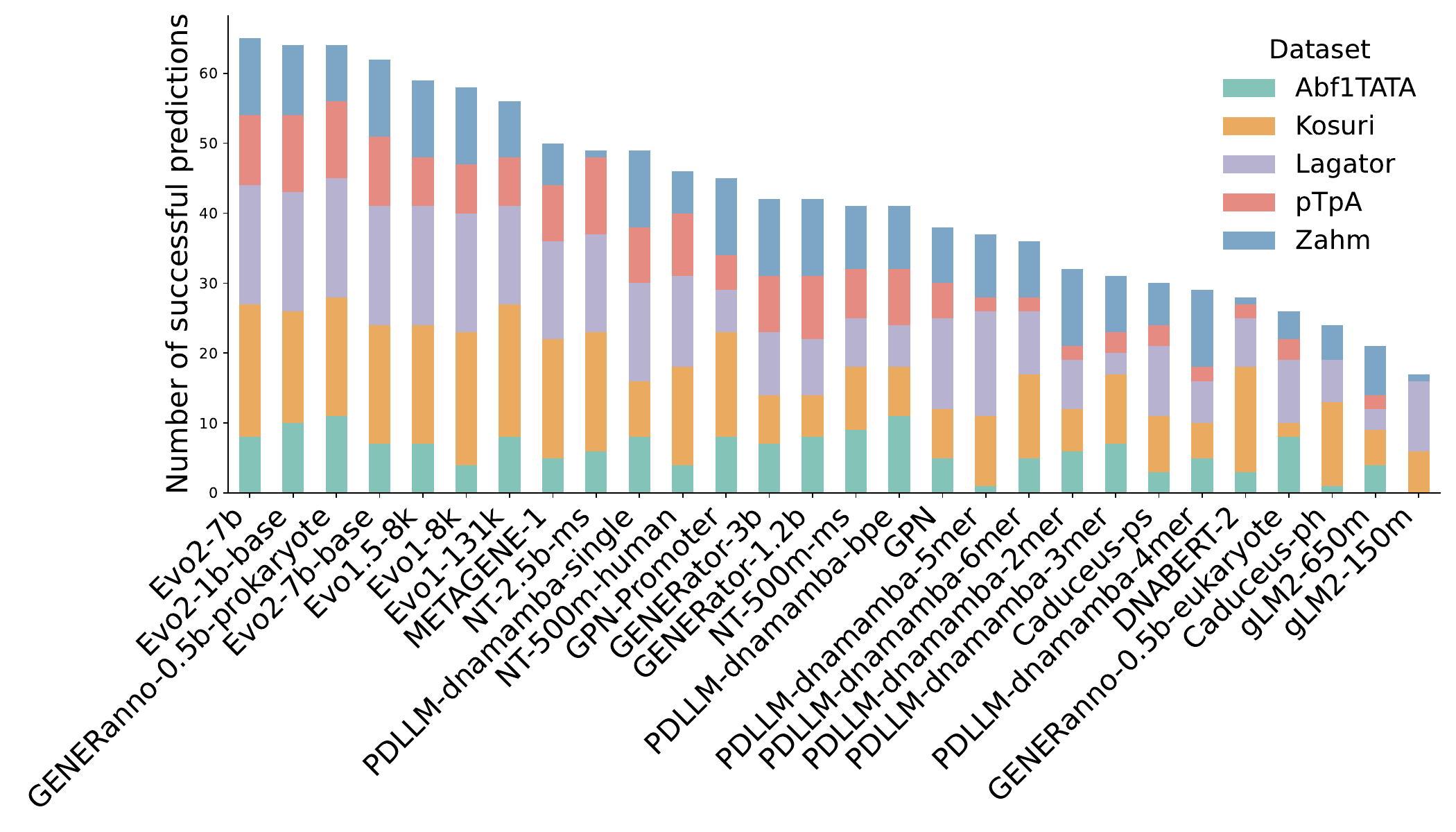}
    \caption{
        \textbf{The number of cumulative successful predictions made by each model series across four datasets: Abf1TATA and pTpA (from deBoer), Zahm, Kosuri, and Lagator.}
    }
    \label{SIfig:fig2}
\end{figure*}

\clearpage

\begin{figure*}[!ht]
    \centering
    \includegraphics[width=0.9\linewidth]{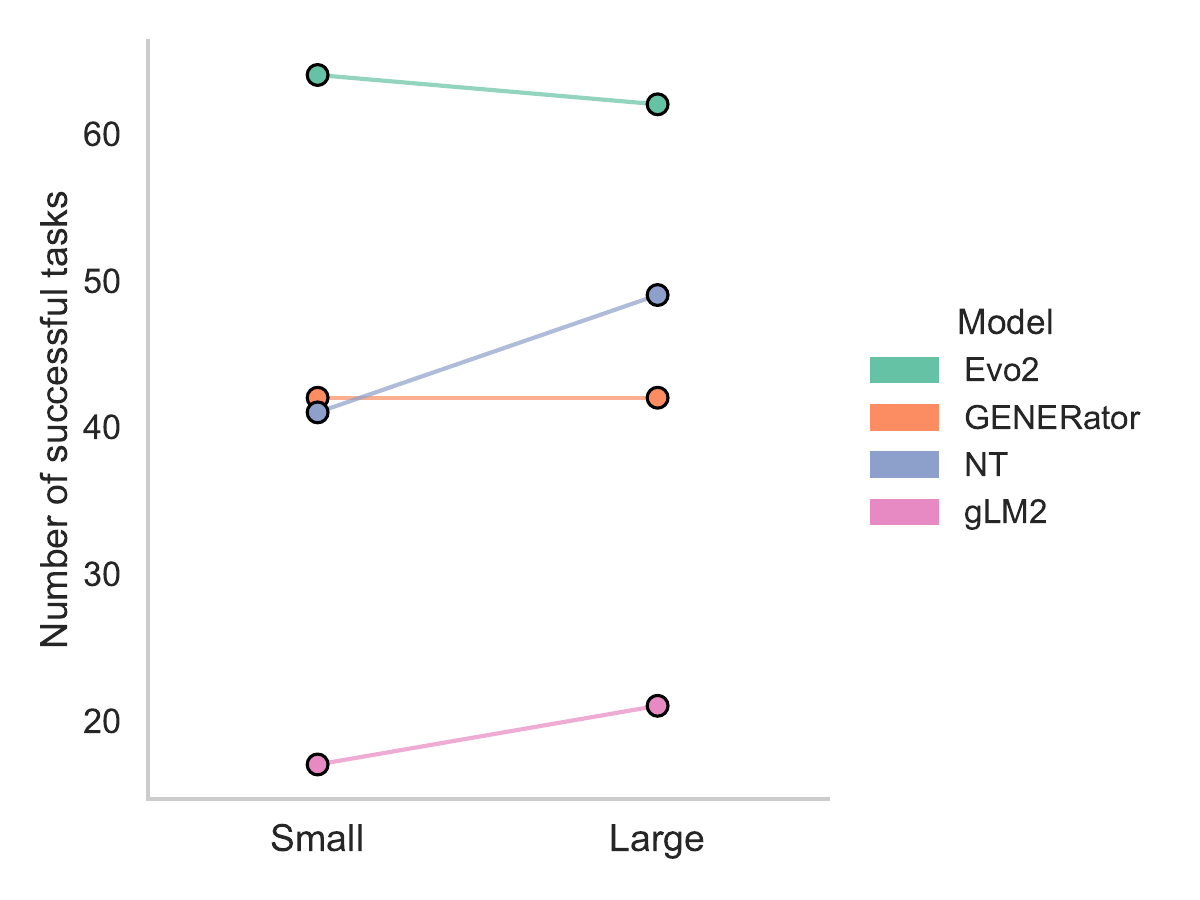}
    \caption{
        Number of Nullsette mutations successfully identified for gLMs with small and large variants, including Evo2 (1B vs. 7B), GENERator (1.2B vs. 3B), NT (500M vs. 2.5B, multispecies), and gLM2 (150M vs. 650M).
    }
    \label{SIfig:fig3}
\end{figure*}




\clearpage
\subsection*{Supplemental Tables}
\setcounter{table}{0} 
\renewcommand{\thetable}{S\arabic{table}} 


\begin{table}[!ht]
\centering
\begin{tabular}{|p{3cm}|p{11cm}|}
\hline
Mutant ID & Description \\
\hline
Translocation-1 & CDS - Promoter - RBS - Start codon - Stop codon - Terminator \\
\hline
Translocation-2 & Promoter - CDS - RBS - Start codon - Stop codon - Terminator \\
\hline
Translocation-3 & Promoter - RBS - CDS - Start codon - Stop codon - Terminator \\
\hline
Translocation-4 & Promoter - RBS - CDS - Stop codon - Start codon - Terminator \\
\hline
Translocation-5 & Promoter - RBS - CDS - Stop codon - Terminator - Start codon \\
\hline
Translocation-6 & Promoter - RBS - Start codon - Stop codon - CDS - Terminator \\
\hline
Translocation-7 & Promoter - RBS - Start codon - Stop codon - Terminator - CDS \\
\hline
Translocation-8 & Promoter - RBS - Start codon - Terminator - CDS - Stop codon \\
\hline
Translocation-9 & Promoter - RBS - Terminator - Start codon - CDS - Stop codon \\
\hline
Translocation-10 & Promoter - Start codon - CDS - RBS - Stop codon - Terminator \\
\hline
Translocation-11 & Promoter - Start codon - CDS - Stop codon - RBS - Terminator \\
\hline
Translocation-12 & Promoter - Start codon - CDS - Stop codon - Terminator - RBS \\
\hline
Translocation-13 & Promoter - Start codon - RBS - CDS - Stop codon - Terminator \\
\hline
Translocation-14 & Promoter - Terminator - RBS - Start codon - CDS - Stop codon \\
\hline
Translocation-15 & RBS - Promoter - Start codon - CDS - Stop codon - Terminator \\
\hline
Translocation-16 & RBS - Start codon - CDS - Promoter - Stop codon - Terminator \\
\hline
Translocation-17 & RBS - Start codon - CDS - Stop codon - Promoter - Terminator \\
\hline
Translocation-18 & RBS - Start codon - Promoter - CDS - Stop codon - Terminator \\
\hline
Translocation-19 & Start codon - Promoter - RBS - CDS - Stop codon - Terminator \\
\hline
\end{tabular}
\caption{\textbf{Prokaryotic virtual mutant cases.} 
Translocation-3,4,5 cannot be compensated by a later start codon as there are no other in-frame start codon.
}
\label{tab:prok_virtual_mutant}
\end{table}

\begin{table}[!ht]
\centering
\begin{tabular}{|p{3cm}|p{11cm}|}
\hline
Mutant ID & Description \\
\hline
Translocation-1 & CDS - Promoter - Start codon - Stop codon - Terminator \\
\hline
Translocation-2 & Promoter - CDS - Start codon - Stop codon - Terminator \\
\hline
Translocation-3 & Promoter - CDS - Stop codon - Start codon - Terminator \\
\hline
Translocation-4 & Promoter - CDS - Stop codon - Terminator - Start codon \\
\hline
Translocation-5 & Promoter - Start codon - Stop codon - CDS - Terminator \\
\hline
Translocation-6 & Promoter - Start codon - Stop codon - Terminator - CDS \\
\hline
Translocation-7 & Promoter - Start codon - Terminator - CDS - Stop codon \\
\hline
Translocation-8 & Promoter - Terminator - Start codon - CDS - Stop codon \\
\hline
Translocation-9 & Start codon - CDS - Promoter - Stop codon - Terminator \\
\hline
Translocation-10 & Start codon - CDS - Stop codon - Promoter - Terminator \\
\hline
Translocation-11 & Start codon - Promoter - CDS - Stop codon - Terminator \\
\hline
\end{tabular}
\caption{\textbf{Eukaryotic virtual mutant cases.} 
Translocation-2,3,4 cannot be compensated by a later start codon as there are no other in-frame start codon.
}
\label{tab:euk_virtual_mutant}
\end{table}

\begin{table}[!ht]
\small
\centering
\begin{tabular}{|p{5cm}|p{10cm}|}
\hline
Model series &  Representative variant \\
\hline
METAGENE1~\cite{METAGENE_1} & METAGENE-1 (METAGENE1) \\
\hline
Nucleotide Transformer~\cite{NT} & nucleotide-transformer-2.5b-multi-species (NT-2.5B-NT) \\
\hline
GENERator~\cite{GENERator} & GENERator-eukaryote-3b-base (GENERator-3B) \\
\hline
GENERanno~\cite{GENERanno} & GENERanno-prokaryote-0.5b-base (GENERanno-0.5B) \\
\hline
Evo1~\cite{Evo1,Evo1_5} & evo-1.5-8k-base (Evo1.5) \\
\hline
Evo2~\cite{Evo2} & evo2\_7b\_base (Evo2-7B) \\
\hline
DNABERT2~\cite{DNABERT2} & DNABERT-2-117M (DNABERT2) \\
\hline
Caduceus~\cite{Caduceus} & caduceus-ph\_seqlen-131k\_d\_model-256\_n\_layer-16 (Caduceus) \\
\hline
GPN~\cite{GPN} & gpn-brassicales (GPN) \\
\hline
GPN-Promoter~\cite{GPNpromoter} & gpn-animal-promoter (GPN-Promoter) \\
\hline
gLM2~\cite{gLM2} & gLM2\_650M (gLM2) \\
\hline
PDLLM~\cite{PDLLM} & PDLLM-dnamamba-6mer (PDLLM) \\
\hline
\end{tabular}
\caption{\textbf{Representative model variant.} For each model series, we list a representative variant used in major benchmarking. The official model identifier is listed, followed in parentheses by the shorthand name used throughout the main text. For example, the variant “nucleotide-transformer-2.5b-multi-species” is referred to as “NT-2.5B-NT” in the text.}
\label{tab:rep_model}
\end{table}

\newpage
\begin{sidewaystable}
\small
\centering
\begin{tabular}{|p{3.6cm}|p{8cm}|p{2cm}|p{3cm}|p{2.3cm}|p{4.5cm}|}
\hline
\small
\textbf{Model name} & \textbf{Pretraining dataset} & \textbf{Pretraining method} & \textbf{Tokenization} & \textbf{Architecture} & \textbf{Input length} \\
\hline
Evo1 8k~\cite{Evo1} & OpenGenome: Prokaryotic whole-genomes dataset (~300B tokens). & CLM & Single-nucleotide tokens & StripedHyena, 7B params & Pretrained with 8,192 context \\
\hline
Evo1 131k~\cite{Evo1} & OpenGenome: Prokaryotic whole-genomes dataset (~300B tokens). & CLM & Single-nucleotide tokens & StripedHyena, 7B params & Pretrained with 131,072 context using Evo1 8k as the base model \\
\hline
Evo1.5~\cite{Evo1_5} & 50\% increase in training data comparing to Evo1 & CLM & Single-nucleotide tokens & StripedHyena, 7B params & Pretrained with 8,192 context using Evo1 8k as the base model \\
\hline
Evo2 1B base~\cite{Evo2} & OpenGenome2: a dataset containing 8.8 trillion tokens from all domains of life. & CLM & Single-nucleotide tokens & StripedHyena 2, 1B params & Pretrained with 8192 context length \\
\hline
Evo2 7B base~\cite{Evo2} & OpenGenome2: a dataset containing 8.8 trillion tokens from all domains of life. & CLM & Single-nucleotide tokens & StripedHyena 2, 7B params & Pretrained with 8192 context length \\
\hline
Evo2 7B~\cite{Evo2} & OpenGenome2: a dataset containing 8.8 trillion tokens from all domains of life. & CLM & Single-nucleotide tokens & StripedHyena 2, 7B params & Pretrained with 1M context using  Evo2 7b base as the base model \\
\hline
GENERator eukaryote 3B~\cite{GENERator} & Eukaryotic genomes (~386B bp), plants, fungi, protozoa, mammlian, vertebrate, invertebrate. & CLM & 6-mer & Llama-based decoder, 3B params & Pretrained with a context length of 98k bp \\
\hline
GENERator eukaryote 1.2B~\cite{GENERator} & Eukaryotic genomes (~386B bp), plants, fungi, protozoa, mammlian, vertebrate, invertebrate. & CLM & 6-mer & Llama-based decoder, 1.2B params & Pretrained with a context length of 98k bp \\
\hline
METAGENE-1 \cite{METAGENE_1} & 1.5T base pairs of DNA and RNA sequences from human wastewater samples. & CLM & Byte-pair encoding & Llama-2, 7B params & Pretrained with 512 sequence length \\
\hline
PDLLM-DNAMamba \cite{PDLLM} & Plant reference genomes & CLM  & 6-mer/5-mer/4-mer/3-mer/2-mer/single/BPE  & SSM, 130M params  &  Pretrained max token length is 512 \\
\hline
\end{tabular}
\caption{Casual language modeling (CLM) based genomic language models}
\label{glm_detail_clm}
\end{sidewaystable}

\begin{sidewaystable}
\small
\centering
\begin{tabular}{|p{3.6cm}|p{8cm}|p{2cm}|p{3cm}|p{2.3cm}|p{4.5cm}|}
\hline
\small
\textbf{Model name} & \textbf{Pretraining dataset} & \textbf{Pretraining method} & \textbf{Tokenization} & \textbf{Architecture} & \textbf{Input length} \\
\hline
Nucleotide Transformer 2.5B multispecies \cite{NT} & 850 whole genomes from NCBI, plants and viruses are not included, resulting in a total of 174B nucleotides, i.e. roughly 29B tokens. & MLM & 6-mers & Transformer, 2.5B params & 1000 bp \\
\hline
Nucleotide Transformer 500M multispecies \cite{NT} & 850 whole genomes from NCBI, plants and viruses are not included, resulting in a total of 174B nucleotides, i.e. roughly 29B tokens. & MLM & 6-mers & Transformer, 500M params & 1000 bp \\
\hline
Nucleotide Transformer 500M Human ref \cite{NT} & GRCh38 human reference genome, resulting in 3B nucleotides, i.e. roughly 500M 6-mer tokens. & MLM & 6-mers & Transformer, 500M params & 1000 bp \\
\hline
GENERanno prokaryote 0.5B base~\cite{GENERanno} & 715 billion base pairs of prokaryotic DNA & MLM & single-nucleotide & Transformer encoder, 500M params & 8k bp \\
\hline
GENERanno eukaryote 0.5B base~\cite{GENERanno} & 386 billion base pairs of eukaryotic DNA & MLM & single-nucleotide & Transformer encoder, 500M params & 8k bp \\
\hline
Caduceus-ph \cite{Caduceus} & Human reference genome (35M tokens / nucleotide base pairs). & MLM & Single-nucleotide tokens & BiMamba, 7.7M params & Pretrained with 131,072 sequence length (reverse complement augmentation) \\
\hline
Caduceus-ps \cite{Caduceus} & Human reference genome (35M tokens / nucleotide base pairs). & MLM & Single-nucleotide tokens & BiMamba, 7.7M params & Pretrained with 131,072 sequence length (No reverse complement augmentation) \\
\hline
GPN \cite{GPN} & Brassicales reference genome from NCBI Genome. Took the union of exons (with a small intronic flank), promoters (1,000 bp upstream of transcription start sites) as well as an equivalent amount of random windows from the whole genome. & MLM & Single-nucleotide tokens & CNN & 512 bp \\
\hline
GPN-Promoter \cite{GPNpromoter} & Animal promoter, genomes of 434 animal species. & MLM & Single-nucleotide tokens & ByteNet, 152M & Pretrained on 512 bp sequences centered at TSSs of protein-coding genes \\
\hline
DNABERT2 \cite{DNABERT2} & Human genome dataset (2.75B nucleotide bases) + Multispecies genome dataset (from 135 species, spread across 6 cetogories). The dataset includes 32.49B nucleotides bases, excluding all sequence with N and retain only ATCG. & MLM & Byte-pair encoding & BERT, 117M params & Pretrained on 700 bp length sequences \\
\hline
gLM2 \cite{gLM2} & OMG: encodes genomic scaffolds with both amino-acid (CDS) and DNA tokens (315B tokens). & MLM & Char-level tokens (DNA: lowercase, AA: uppercase) & Transformer, 650M params / 150M params & Pretrained with a 4096 token context window \\
\hline
\end{tabular}
\caption{Mask language modeling (MLM) based genomic language models}
\label{glm_detail_mlm}
\end{sidewaystable}


\clearpage

\end{document}